\documentclass[aps,twocolumn,showpacs]{revtex4}
\usepackage{graphicx}
\usepackage{amsmath}
\usepackage{amssymb}

\def\bc{\begin{center}}
\def\ec{\end{center}}

\def\beq{\begin{equation}}
\def\eeq{\end{equation}}

\begin{document}

\title[Ising instability of a phonon mode in graphene]
{ Ising instability of a Holstein phonon mode in graphene }

\author{K. Ziegler$^1$, E. Kogan$^2$, E. Majern\'ikov\'a$^3$ and S. Shpyrko$^4$}
\address{$^1$ Institut f\"ur Physik, Universit\"at Augsburg, 86135 Augsburg, Germany\\
$^2$ Department of Physics, Bar--Ilan University, Ramat Gan 52900, Israel\\
$^3$Institute  of Physics, Slovak Academy of Sciences, D\'ubravsk\'a cesta 9, SK-84511 Bratislava, Slovak Republic\\
$^4$ Institute for Nuclear Research, Ukrainian Academy of Sciences, pr. Nauki 47, Kiev, Ukraine
}

\date{\today}

\begin{abstract}
We study the thermal distribution of phonons in a graphene sheet.
Due to the two electronic bands there are two out-of-plane 
phonon modes with respect to the two sublattices. 
One of these modes undergoes an Ising transition by spontaneously breaking the
sublattice symmetry. We calculate the critical point, the renormalization of the
phonon frequency and the average lattice distortion. This transition might be 
observable in Raman scattering and in transport properties.
\end{abstract}


\maketitle


Transport measurements have revealed many interesting properties of the two-dimensional
material graphene \cite{Novoselov2005,zhang05,Novoselov2006,Katsnelson2006,C,Abergela:2010}.
More recently, Raman scattering has become an additional experimental method to study
physical properties of graphene \cite{ferrari06,pisana07,elias09}. 
It provides a deeper insight into the elastic properties of
the honeycomb lattice of carbon atoms and the related electron-phonon interaction.
Since the latter is important also for some of the unusual transport
properties, these results are of great value for a better understanding of graphene.

Recent experiments on hydrogenated graphene (graphane) have revealed additional Raman 
spectral lines due to hydrogen atoms \cite{elias09}, indicating that there is an important 
electron-phonon interaction between the conduction electrons and the lattice formed by the 
hydrogen atoms. These results indicate that doping can alter the properties of graphene
by changing the band structure as well as by modifiying the elastic properties due to
phonons. 

Theoretical studies of electron-phonon interaction in graphene at zero temperature came 
to a mixed conclusion. Although the electron-phonon coupling is remarkably strong in terms
of a renormalization of the lattice vibrations, its effect 
on transport properties is rather weak \cite{castro07,C}. In particular, the out-of-plane 
optical mode (ZO phonon) has almost no effect on transport at all \cite{stauber08}. 
Some experiments also provide evidence that transport properties are not much affected 
by phonons \cite{morozov08}. This requires a better understanding of the electron-phonon 
interaction and its consequences for the transport properties.

In this paper we will explain that the ZO phonon mode can play an important role by 
spontaneously breaking the chiral (sublattice) symmetry of graphene. This effect is
accompanied by an Ising transition which is associated with a softening of the optical
phonon mode.


{\it model:} The dynamics of phonons of a honeycomb lattice is complex and can be described by
a number of phonon modes \cite{falkovsky07,castro07,stauber08,C,sasaki08}. For simplicity, 
we focus here on optical phonons, coupled to the electrons by a Holstein interaction 
\cite{stauber08}.
Then in a tight-binding description, electrons coupled to optical (Einstein) phonons 
at frequency $\omega_0$ are described by the Holstein Hamiltonian as
\beq
H= \omega_0 \sum_{\bf r} b_{\bf r}^\dagger b_{\bf r} + \sum_{{\bf r},{\bf r}'}h_{{\bf r},{\bf r}'}
c^\dagger_{\bf r} c_{{\bf r}'}
+\alpha\sum_{\bf r}(b_{\bf r}+b^\dagger_{\bf r})c^\dagger_{\bf r} c_{\bf r}
\ .
\label{hamilton00}
\eeq
Here $c^\dagger_{\bf r}$ ($c_{\bf r}$) are the electron creation (annihilation) operators and 
$b^\dagger_{\bf r}$ ($b_{\bf r}$) are the phonon creation (annihilation) operators.
Then the unitary Lang-Firsov transformation \cite{lang63} reveals that this Hamiltonian 
describes an effective attraction between the fermions and a renormalization of the 
electronic hopping amplitude $h_{{\bf r},{\bf r}'}$, 
reducing the hopping rate substantially (polaron effect). The attractive interaction 
competes with the repulsive Coulomb interaction, where the latter might be dominant.

The conventional approach to determine the properties of a coupled system of
phonons and electrons is based on
a self-consistent evaluation of the self energy (Migdal approximation) 
\cite{park07,deveraux07}. The latter provides an effective (or renormalized)
energy and its imaginary part an effective scattering rate. Such a static approximation might be 
insufficient in a two-dimensional system, since it does not take into account thermal fluctuations. 
This was already discussed in an experimental study of graphene \cite{pisana07}. 
To avoid this problem, we include thermal fluctuations in our approach.
To this end, we replace the phonon operators $b_{\bf r}$, $b^\dagger_{\bf r}$ by their
quantum average: $b_{\bf r}\approx \langle b_{\bf r}\rangle\equiv v_{\bf r}$ and 
$b_{\bf r}^\dagger\approx \langle b^\dagger_{\bf r}\rangle\equiv v_{\bf r}^*$. In this 
approximation we can keep thermal fluctuations but ignore quantum fluctuations of the phonons.
The electrons, on the other hand, are studied in full quantum dynamics. This reduces the
grand-canonical ensemble at inverse temperature $\beta$
to a functional integral with respect to thermal fluctuations of the lattice distortions
$v_{\bf r}$ and a trace with respect to the electrons. It should be noticed that only 
the real part of the phonon field $v$ couples to the electrons.
After performing the trace over the electrons
we get the partition function 
\beq
Tr e^{-\beta H}
\approx\int \det({\bf 1}+e^{-\beta h})e^{-\beta S_0}   
{\cal D}[v]\equiv Z
\label{genfunc1}
\eeq
with the phonon dispersion 
\beq
S_0=\frac{\omega_0}{2}\sum_{\bf r}{\vec v}_{\bf r}^2 
\ .
\label{phonons}
\eeq

It is convenient to introduce a sublattice representation for the tight-binding Hamiltonian $h$,
since the graphene unit cell contains two atoms, each of them has one $\pi$-orbital.
This gives a two-component wavefunction, and the electronic Hamiltonian can be expressed
by $2\times2$ (Pauli) matrices as
\beq
h_e=
\begin{pmatrix}
-\mu & {\hat t} \\
{\hat t}^T & -\mu,
\end{pmatrix}
\equiv h_1\sigma_1+h_2\sigma_2-\mu\sigma_0
\ .
\label{hamiltonian0}
\eeq
with $h_1=({\hat t}+{\hat t}^T)/2$, $h_2=i({\hat t}-{\hat t}^T)/2$ and with
the chemical potential (bare Fermi energy) $\mu$. In case of the honeycomb lattice
${\hat t}$ is a matrix that connects nearest-neighbor sites on the lattice
\[
{\hat t}_{{\bf r},{\bf r}'}=t\sum_{j=1}^3\delta_{{\bf r}',{\bf r}+{\bf a}_j}
\ ,
\]
where ${\bf r}$ is on sublattice A and ${\bf r}+{\bf a}_j$ is on sublattice B.

The electron-phonon coupling reads in this sublattice representation as (spatial
dependence of the distortions are implictly assumed)
\[
h_{e-ph}=\alpha
\begin{pmatrix}
v_A & v_1-iv_2\\
v_1+iv_2 & v_B 
\end{pmatrix}
\]
\[
=
\psi\sigma_0+\phi\sigma_3+\alpha(v_1\sigma_1+v_2\sigma_2)
\ ,
\]
where $\psi=\alpha(v_A+v_B)/2$ and $\phi=\alpha(v_A-v_B)/2$.
The in-plane phonon modes $v_1$, $v_2$ will be ignored in the following because they preserve 
the chiral symmetry of the electronic part of the Hamiltonian. Consequently, they are not able 
to cause symmetry breaking. More interesting are the out-of-plane modes $\psi$ and 
$\phi$ because they break the chiral symmetry of $h_e$.
Then the phonon term of Eq. (\ref{phonons}) becomes
\[
S_0=\frac{\omega_0}{\alpha^2}\sum_{\bf r}\left(\psi_{{\bf r}}^2 +\phi_{{\bf r}}^2\right)
\ .
\]
We introduce the rescaled frequency $\omega=\omega_0/\alpha^2$. 
Eventually, the combined Hamiltonian reads
\beq
h=h_e+h_{e-ph}=h_1\sigma_1+h_2\sigma_2   
+(\psi-\mu)\sigma_0+\phi\sigma_3 
\ .
\label{hamiltonian1}
\eeq
$Z$ of Eq. (\ref{genfunc1}) serves as a generating function that allows us to get, 
for instance, the static electronic Green's by differentiation of $\ln Z$ as
\beq
G_{{\bf r},{\bf r}'}=\frac{1}{Z}\int ({\bf 1}+e^{-\beta h})^{-1}_{{\bf r},{\bf r}'}
e^{-\beta S}{\cal D}[{\phi,\psi}]
\label{fint}
\eeq
with $S=S_0-\beta^{-1}\ln\det({\bf 1}+e^{-\beta h})$.
It is important to notice that $e^{-\beta S}=e^{-\beta S_0}\det({\bf 1}+e^{-\beta h})$ 
is a non-negative function. Therefore, $e^{-\beta S}/Z$ is a probability
density for the phonon field, and the static one-particle Green's function then 
can be rewritten as an average $\langle ... \rangle_{ph}$ with respect to 
the distribution $e^{-\beta S}/Z$ \cite{ziegler05}:
\beq
G_{{\bf r},{\bf r}'}=\langle ({\bf 1}+e^{-\beta h})^{-1}_{{\bf r},{\bf r}'} \rangle_{ph}
\ .
\label{average_green}
\eeq
Without electron-phonon interaction the distribution of lattice vibrations is
$\exp(-\beta S_0)$ such that the average lattice distortion vanishes: 
$\langle \phi\rangle_{ph}=\langle \psi\rangle_{ph}=0$. A nonzero $\langle \psi\rangle_{ph}$ 
presents a shift of the Fermi energy, whereas a nonzero $\langle\phi\rangle_{ph}$ would break 
the sublattice symmetry.
An interesting question is whether or not a coupling to electrons can create a 
nonzero average distortion. From the symmetry point of view this should not be the
case for $\phi$ because the system is invariant under the transformation
$\phi\to-\phi$. This is a consequence of the fact that
the distribution $e^{-\beta S}/Z$ 
is invariant under unitary transformations. Choosing the discrete unitary 
transformation $U=(\sigma_1+\sigma_2)/\sqrt{2}$ we get
\beq
UhU^\dagger=h_1\sigma_2+h_2\sigma_1+ (\psi-\mu)\sigma_0-\phi\sigma_3
\ .
\label{symm_trans}
\eeq
This means for $h$ in Eq. (\ref{hamiltonian1}) an exchange of $h_1$ with $h_2$ 
and a sign change $\phi\to-\phi$.
Since the system is assumed to be isotropic, only a change $\phi\to-\phi$ remains. 
This is the invariance under a $Z_2$ (Ising-like) transformation.
Thus $\langle\phi\rangle_{ph}\ne 0$ represents a typical problem of 
spontaneous symmetry breaking.

\begin{figure}[t]
\begin{center}
\includegraphics[width=8cm,height=6cm]{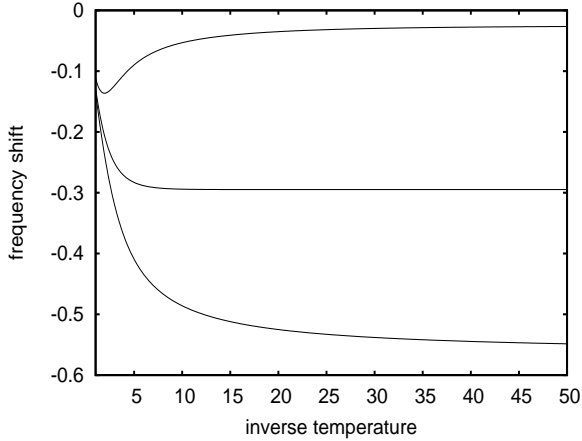}
\caption{Frequency shift $\delta\omega_{33}$ in units of $1/t$ vs. inverse 
temperature $\beta t$ for $\mu/t=0, 3, 6$
(from bottom to top) from Eq. (\ref{second_der0}).}
\label{fig:renorm}
\end{center}
\end{figure}

In previous calculations only the $\psi$ mode was taken into account, whereas the $\phi$ mode 
has been neglected by assuming $\phi=0$ \cite{castro07,stauber08}. Indeed, a solution with 
$\phi=0$ always exists due to the symmetry in Eq. (\ref{symm_trans}). However, spontaneous 
symmetry breaking is possible. This would be associated with an instability of the $\phi=0$ 
solution. Although the symmetry-breaking solution is not available from a simple perturbation 
theory, the latter can be employed to analyze the stability of the symmetry-preserving solution. 
For this purpose we evaluate the shift of the phonon frequency $\omega_0\delta_{jl}\to
\omega_0\delta_{jl}+\delta \omega_{jl}$ ($j=0$: $\psi$ mode, $j=3$: $\phi$ mode) for ${\bf q}=0$ 
(i.e. at the $\Gamma$ point of the Brillouin zone). In second-order perturbation theory in $\alpha$ 
\cite{castro07,stauber08,C} this gives for our model a diagonal matrix with
$\delta \omega_{00}\sim-const./\beta$ and
\beq
\delta \omega_{33}
=-\frac{\alpha^2}{2\pi\beta t^2}
\ln\left[
\frac{\cosh(\beta t\Lambda)+\cosh(\beta\mu)}{1+\cosh(\beta\mu)}
\right] 
\ ,
\label{second_der0}
\eeq
where $\Lambda=2\sqrt{\pi}$ is the momentum cutoff.
Thus the renormalization of $\omega_0$ of the $\psi$ mode vanishes at low temperatures. 
The situation is different for the $\phi$ mode. Although $\delta \omega_{33}$ also
decreases with decreasing temperature, it has a non-zero value in the limit of zero 
temperature (cf. Fig. \ref{fig:renorm}). Then the solution $\phi=0$ is stable (unstable) 
if $\omega_0+\delta \omega_{33}\ge0$ ($<0$). An unstable situation with negative 
renormalized phonon frequency of a different mode was also found by Castro Neto et al. 
\cite{C}.

In order to find a stable solution for the case of $\omega_0+\delta \omega_{33}<0$
we must go beyond perturbation theory. One possibility is to extend the second-order
calculation by a partial resummation of the perturbation series to all orders of $\alpha$. 
This would provide a self-consistent calculation for the phonon modes. 
However, it is more convenient here to use a saddle-point (SP) approximation for the 
$\phi$, $\psi$ integration in Eqs. (\ref{genfunc1}), ({\ref{fint}). This calculation 
can be extended to include fluctuations around the SP solution. 
The latter allows us to analyze the stability of the SP solution, 
to identify possible instabilities and to determine the phonon dispersion.


The SP approximation is based on the variational equation $\delta S=0$:
\beq
\beta\omega\phi_j = \frac{\partial}{\partial\phi_j}\ln\left[
\det({\bf 1}+e^{-\beta h})\right]
\ \ \ (j=0,3)
\label{SPE}
\eeq
with $\phi_0=\psi$ and $\phi_3=\phi$.
Now we assume that the SP solution $\psi,\phi$ is uniform in space and time.
Thus our SP solution is a mean-field approximation.
This allows us to diagonalize the argument of the trace term in Eq. (\ref{SPE})
using a Fourier representation. Moreover, we approximate the Hamiltonian
$h_e$ by its low-energy Dirac behavior $h_e\sim k_1\sigma_1+k_2\sigma_2$ now. 
Then the eigenvalues of $h$ are
$
\lambda_{\pm}= \psi-\mu\pm \sqrt{\phi^2+t^2k^2} \label{ef}
$ and the SP equations read
\[
\omega\psi = -1 -\frac{1 }{2\pi}
\]
\beq
\times \int\limits_0^{\Lambda}\frac{ \sinh(\beta(\mu- \psi))}{\cosh(\beta
  (\mu-\psi))+\cosh(\beta\sqrt{\phi^2+t^2k^2})}k dk
\label{psi1}
\eeq
\[
\omega\phi = \int\limits_0^{\Lambda} \frac{\phi}
  {\sqrt{\phi^2+t^2k^2}}
\]
\beq
\times\frac{\sinh(\beta\sqrt{\phi^2+t^2k^2})}
  {\cosh(\beta(\mu-\psi))+\cosh(
  \beta\sqrt{\phi^2+t^2k^2})}\frac{k dk}{2\pi}
\ .
\label{phi1}
\eeq
The integration of the second equation can be performed and gives
\beq
\omega \phi = \frac{1}{2\pi\beta t^2}\phi \ln\left[\frac{\cosh(\beta\sqrt{\phi^2
  +t^2\Lambda^2})+
  \cosh(\beta {\bar\mu})}{\cosh(\beta\phi)+ \cosh(\beta{\bar\mu})}\right]
\label{phi2}
\eeq
with the renormalized chemical potential ${\bar\mu}=\mu-\psi$. 

If the mean-field solution $\phi=0$ is unstable 
we have to find a solution $\phi\ne 0$ of Eq. (\ref{phi2}).
This can be obtained from a quadratic equation after an expansion of 
Eqs. (\ref{psi1}), (\ref{phi2}) for small $\phi$ and ${\bar\mu}$. Then the equation for $\phi$ reads
\begin{equation}
\label{Phi}
\frac{\partial S}{\partial\phi}=
\Omega\phi+\Gamma \phi^3 +C{\bar\mu}^2\phi=0
\end{equation}
with the coefficients
\[
\Omega=
\omega -\frac{1 }{\pi\beta t^2}\ln\cosh \left(\beta t \sqrt{\pi}\right )
\]
\[
C=\frac{\beta\tau^2}{8\pi t^2} , \ \ \
\Gamma=\frac{\beta}{4\pi t^2}\left(1-\frac{\tau 
}{\beta t\sqrt{\pi}}\right)
\]
with $\tau=\tanh (\beta t \sqrt{\pi})$. 
Besides the vanishing mean-field solution $\phi=0$ there are now the two non-vanishing solutions
\begin{equation} \phi_\pm=\pm
\frac{1}{\sqrt{\Gamma}}\sqrt{-\Omega-C{\bar\mu}^2}
\ .
\label{solution1}
\end{equation}
The power law with exponent $1/2$ is the result of the mean-field approximation.
Due to the Ising-like symmetry of Eq. (\ref{symm_trans}),
the exponent should be $1/8$ instead of $1/2$ \cite{mccoy}. However, this incorrectness
may only be important very close to the transition point. 

Moreover, in a small vicinity around the Dirac point ${\bar\mu}=0$ the SP equation (\ref{psi1}) 
can be linearized. For small $\phi$ this leads to a renormalized Fermi energy
\beq
{\bar\mu}\approx \frac{\mu+1/\omega}{1-\ln (2)/\pi\beta \omega t^2}
\ .
\label{chem_pot2}
\eeq

According to the discussion of Eq. (\ref{second_der0}), a mean-field solution $\phi=0$ is unstable when
$\Omega+C{\bar\mu}^2<0$.  
For $\beta t\gg 1$ we have
\beq
\Omega+C{\bar\mu}^2\sim \omega-\frac{1}{\sqrt{\pi}t}+\frac{\ln 2}{\pi\beta t^2}
+\frac{\beta{\bar\mu}^2}{8\pi t^2}
\ .
\label{stability2}
\eeq
It should be noticed here that with decreasing temperature $\Omega + C{\bar\mu}^2$ {\it decreases}
at the Dirac point ${\bar\mu}=0$ but it {\it increases} sufficiently away from the Dirac point. 
This is similar to the behavior shown in Fig. \ref{fig:renorm}.
In particular, for $\omega<1/\sqrt{\pi}t$ and $\beta t\sim \infty$ we have $\phi=0$ if ${\bar\mu}\ne0$ 
and $\phi\ne0$ for ${\bar\mu}=0$.
Thus, at sufficiently low temperatures a symmetry-broken mean-field solution exists only at
the Dirac point. Away from the Dirac point there is no symmetry breaking at very low temperatures.

For the case $\Omega+C{\bar\mu}^2<0$ we must replace $\phi=0$ by one of the solutions of Eq. (\ref{solution1}).
Inserting the proper solution into the expression of the renormalized phonon frequency 
$\partial^2 S/\partial\phi^2$, we obtain the non-negative value for the $\phi$ phonon mode
\beq
\omega_\phi=\alpha^2\frac{\partial^2 S}{\partial\phi^2}
\approx \alpha^2|\Omega+C{\bar\mu}^2|\left[1+\Theta(-\Omega-C{\bar\mu}^2)\right]
\ ,
\label{ren_frequency}
\eeq
where $\Theta(...)$ is the Heaviside function. The behavior of the renormalized
phonon frequency is depicted in Fig. \ref{fig:transition}. This frequency vanishes
as a function of the renormalized Fermi energy at the critical point 
$\mu_c=\sqrt{-\Omega/C}$, provided that $\Omega<0$.
For ${\bar\mu}^2<\mu_c^2$ we have spontaneous symmetry breaking with $\langle\Phi\rangle_{ph}\ne 0$
and for ${\bar\mu}^2\ge \mu_c^2$ a symmetric phase with $\langle\phi\rangle_{ph}=0$.
The softening of the phonons at the critical point should be observable in Raman scattering.
Symmetry breaking, on the other hand, might be observable in transport measurements as a metal-insulator
transition.
However, large fluctuations of $\phi$ can destroy the gap even for $\langle\Phi\rangle_{ph}\ne 0$,
similar to the case of a quenched random gap \cite{ziegler09,medvedyeva10,ziegler10}.

The values of the model parameter can be compared with experimental data and bandstructure
calculations. The electronic hopping parameter is $t\approx 2.7$ eV \cite{C} and the bare 
phonon frequency is $\omega_0\approx 0.1$ eV \cite{wirtz04}. Although there is no reliable
estimate for $\alpha$, the parameter $\omega$ might be small enough to have $\Omega<0$, 
which is necessary to see spontaneous symmetry breaking. Even if this is not the case,
$\omega$ can be reduced by doping with non-carbon atoms (e.g. with hydrogen, oxygen or flour).

In conclusion, we have found that in the Holstein model for graphene an out-of-plane phonon mode can
spontaneously break the sublattice symmetry, leading to a spatially fluctuating gap whose mean value
is nonzero.
This effect is accompanied by an Ising-like transition. Although it is not clear whether this
can be observed in pristine graphene, doping with non-carbon atoms will provide conditions 
to observe such a transition. 

\begin{figure}[t]
\begin{center}
\includegraphics[width=8cm,height=6cm]{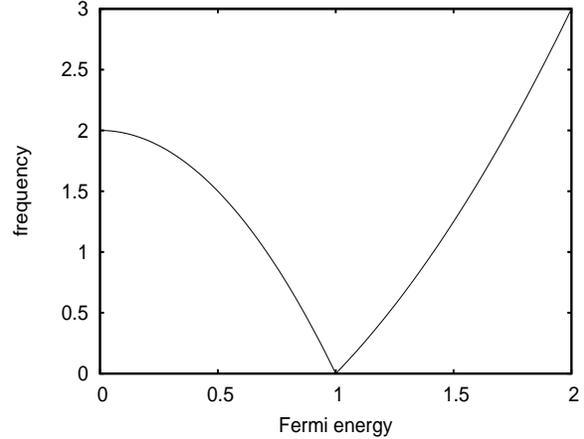}
\caption{Renormalized frequency $\omega_\phi$ of the $\phi$ mode (in arbitrary units) vs. 
Fermi energy ${\bar\mu}$ (in units of $\Omega/C$) from Eq. (\ref{ren_frequency}).}
\label{fig:transition}
\end{center}
\end{figure}


\noindent
{\it acknowledgement:} We acknowledge financial support by the DFG grant ZI 305/5-1
and by the VEGA grant project 2/0095/09.

\end{document}